\documentclass[prb,twocolumn,showpacs,groupedaddress]{revtex4}
\usepackage[dvips]{epsfig}

\begin{document}
\title{Transport properties of moderately disordered UCu$_4$Pd}

\author{A. Otop, S. S\"ullow}
\affiliation{Institut f\"ur Physik der Kondensierten Materie, TU Braunschweig, 38106 Braunschweig, Germany}

\author{M.B. Maple}
\affiliation{Department of Physics and Institute for Pure and Applied Physical Sciences, University of California, San Diego, California 92093, USA}

\author{A. Weber}
\affiliation{EKM, Institut f\"ur Physik, Universit\"at Augsburg, Universit\"atsstra\ss e 1, 86159 Augsburg, Germany}

\author{E.­W. Scheidt}
\affiliation{CPM, Institut f\"ur Physik, Universit\"at Augsburg, Universit\"atsstra\ss e 1, 86159 Augsburg, Germany}

\author{T.J. Gortenmulder}
\affiliation{Kamerlingh Onnes Laboratory, Leiden University, 2300 RA Leiden, The Netherlands}

\author{J.A. Mydosh}
\affiliation{Kamerlingh Onnes Laboratory, Leiden University, 2300 RA Leiden, The Netherlands \\
Max Planck Institute for Chemical Physics of Solids, 01187 Dresden, Germany}

\date{\today}

\begin{abstract}

We present a detailed study on the (magneto)transport properties of as-cast and heat treated material UCu$_4$Pd. We find a pronounced sample dependence of the resistivity $\rho$ of as-cast samples, and reproduce the annealing dependence of $\rho$. In our study of the Hall effect we determine a metallic carrier density for all samples, and a temperature dependence of the Hall constant which is inconsistent with the Skew scattering prediction. The magnetoresistive response is very small and characteristic for spin disorder scattering, suggesting that overall the resistivity is controlled mostly by nonmagnetic scattering processes. We discuss possible sources for the temperature and field dependence of the transport properties, in particular with respect to quantum criticality and electronic localization effects.
\end{abstract}


\maketitle

\section{Introduction}
By now, it is a well-established fact that crystallographic disorder plays a dominant role in the physics of heavy-fermion compounds. There are two major aspects to this issue, namely, the disorder induced suppression of unconventional superconductivity in heavy-fermion superconductors \cite{dalichaouch,sullow1,joynt,zapf}, and secondly the modifications of heavy-fermion normal state properties as consequence of the disorder. The latter aspect has been prominent particularly in the context of quantum critical behavior \cite{miranda,andrade,castro,rosch}, where close to magnetic instabilities both electronic correlations and disorder effects are very much enhanced \cite{wysokinski}.

Experimentally, the relevance of disorder effects for the understanding of heavy-fermion normal state properties has been investigated for various families of compounds. Here, in particular, studies have been performed on pseudo-binary cerium or uranium intermetallics with stoichiometries (Ce,U):$T$ ($T$ = intermetallic element) of 1:1, 1:2 or 1:3 \cite{garca,shlyk,pechev,gajewski}, or ternary uranium heavy fermions like those of composition 2:1:3 \cite{li,nishioka,huo,tran}, the U(Cu,Pd)$_5$ system \cite{bernal,vollmer,aronson,booth}, or U- and Ce-122 compounds \cite{sullow2,sullow3,graf,kalvius,tien}. The goal of such material studies is the search for quantum critical behavior close to a magnetic instability, which might be accessed by tuning a magnetic transition temperature to zero through alloying. The presence of crystallographic disorder complicates matters as it introduces a new, independent, and often uncontrolled parameter.

In a Doniach-like picture, the magnetic phase diagram of crystallographically ordered heavy fermions is thought to be accessible by means of pressure or chemical pressure (through isoelectronic alloying) experiments \cite{doniach,sullow4}. The control parameter to traverse this phase diagram is the magnetic hybridization strength \emph{J}, which in Ce or U compounds increases with pressure. Close to the magnetic instability, it is observed that the magnetic transition temperature continuously is suppressed to zero temperature as function of the control parameter, with $T_N \propto (J_C - J)^x$. At $J_C$, non-Fermi-Liquid (NFL) behavior, i.e., temperature dependencies of the magnetic ground state properties not consistent with Fermi-Liquid theory, often is observed. At $J > J_C$, below a characteristic temperature $T_{FL} \propto (J - J_C)^y$ Fermi-Liquid (FL) behavior is recovered. The exponents $x$ and $y$, in models put forth so far \cite{hertz,millis}, depend on spatial dimensionality and type of magnetic coupling. At present, however, there is no full consensus about the adequacy of these models and the underlying physics \cite{si}.

Recently, the Doniach picture has been extended in order to incorporate disorder effects \cite{theumann,kiselev}. Through the disorder modelled, assuming a distribution $P(J_{ij})$ of locally and randomly varying hybridization values $J_{ij}$, it is demonstrated that the disorder can give rise to spin glass phases, which compete with the other ground states of the Doniach model \cite{doniach}.

In experiments such as alloying studies, aside from chemical pressure effects, the distributions $P(J_{ij})$ are also varied. Therefore, through alloying, two independent control parameters are modified at the same time: (i.) The exerted chemical pressure gives rise to a variation of the average hybridization strength $J$; (ii.) The local level of random site exchange, and thus $P(J_{ij})$, is affected. Qualitatively, $P(J_{ij})$ can be parameterized as a distribution of local hybridization values within a range $\Delta J$ around the average value $J$, i.e., the local hybridization lying in the interval $J \pm \Delta J$. With this terminology, in an alloying study the magnetic ground state properties of a series of materials are a function of the two independent control parameters $J$ and $\Delta J$, with $J$ as a measure for the external pressure and $\Delta J$ for the level of disorder.

\begin{figure}[!ht]
\begin{center}
\includegraphics[width=1\linewidth]{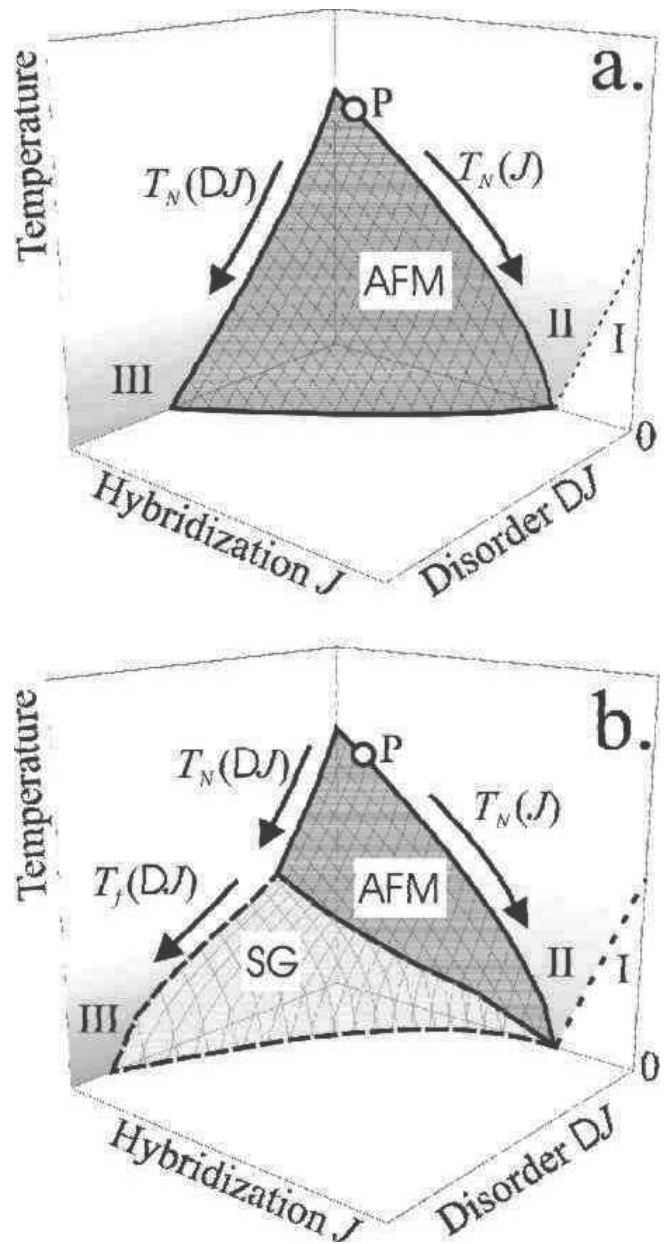}
\end{center}
\caption{Schematic phase diagrams of heavy-fermion compounds as function of the independent control parameters hybridization $J$ and disorder $\Delta J$. For entirely antiferromagnetic coupling there are only paramagnetic and antiferromagnetic (AFM) phases (a). If frustration and competing interactions are taken into account, additionally a spin glass phase (SG) might occur (b). Region I represents a Fermi-Liquid regime, while non-Fermi-Liquid behavior is observed in the regions II and III; for details see text.} \label{fig:fig1}
\end{figure}

The presence of two control parameters $J$ and $\Delta J$ requires an extension of the ''generic magnetic phase diagram'' of heavy fermions by incorporating a third dimension, crystallographic disorder, with $\Delta J$ as parameter characterizing the disorder level. A detailed and quantitative discussion of this ''three dimensional magnetic phase diagram of disordered heavy fermions'' is impossible at present \cite{theumann,kiselev}, and is beyond the scope of this work. A qualitative discussion, however, illustrates various important issues of heavy fermion physics. Therefore, in Fig. \ref{fig:fig1}(a) we sketch the magnetic phase diagram under incorporation of the parameter disorder $\Delta J$.

If we consider a crystallographically perfectly ordered heavy-fermion material, which orders antiferromagnetically (AFM) below a transition temperature $T_N$, in Fig. \ref{fig:fig1}(a) this material would be positioned on the AFM phase transition line for a disorder level $\Delta J = 0$, say, at point $P$. By increasing the hybridization strength $J$, for instance by means of a pressure experiment on a Ce compound, we traverse the ''generic magnetic phase diagram of ordered heavy fermions''. Following Refs. \cite{doniach,millis}, for the three-dimensional case a suppression of the antiferromagnetic transition temperature $T_N \propto (J_C - J)^{2/3}$ is predicted. FL behavior is recovered below $T_{FL} \propto (J - J_C)$ (area ''I''), while at the magnetic instability, for $T_N = 0$, NFL behavior occurs in the shaded area ''II''.

In contrast, again with the starting point of the perfectly ordered material at point $P$, if we increase the level of disorder, it will cause a suppression of the magnetic transition temperature. While there are no detailed predictions on the disorder dependence of $T_N$, qualitatively the suppression has been demonstrated both on theoretical as experimental grounds \cite{griffiths,zhou}. For the low moment magnets and strongly correlated electron systems considered here, because of the inherent instability of the magnetic ground state, a complete suppression of the AFM state through disorder will be possible \cite{note1}. Experimentally, this is indicated by observations on CeAl$_3$ or CePd$_2$Al$_3$ \cite{lapertot,mentink}, where the presence or absence of magnetic order in single or polycrystals has been linked to crystallographic disorder on the Al sites. In consequence, sufficiently close to the magnetic instability (for sufficiently strong correlations) disorder induced suppression of AFM order will occur, leading to a situation as sketched in Fig. \ref{fig:fig1}(a). For $T_N = 0$, NFL behavior is predicted to occur, although now extended over some range in disorder (area III) because of rare magnetic configurations \cite{castro}.

Our phenomenological sketch illustrates two issues relevant to heavy fermion and other strongly correlated electron materials. The first issue concerns how to connect the NFL areas II and III. If there is a continuous and smooth transition from II to III, it would imply that NFL characteristics of a heavy fermion material, even in the presence of weak disorder, could be disorder modified, compared to the perfectly ordered case. Specifically, if an experiment is chosen on an antiferromagnetic and weakly disordered heavy fermion, which only varies the global hybridization strength $J$, upon suppression of $T_N$ its $J$ dependence might deviate from the ordered case, and NFL behavior could occur over a wider range in $J$ even at zero temperature.

Secondly, in alloying experiments both parameters of the phase diagram in Fig. \ref{fig:fig1}(a), hybridization $J$ and disorder $\Delta J$, are modified simultaneously. Correspondingly, in a doping study starting with a structurally perfectly ordered, antiferromagnetic heavy fermion system and finishing at a paramagnetic and structurally ordered Fermi liquid, a complicated trajectory through the three-dimensional phase diagram is traversed. Then, the associated alloying phase diagram will not match an experiment varying J only.

The underlying assumption of the sketched phase diagram in Fig. \ref{fig:fig1}(a) is that upon increasing the disorder level the principal type of magnetic order remains the same. This assumption is not always fulfilled for real materials. If disorder is associated by competing interactions or frustration, it can give rise to new magnetic states, for instance spin glass phases. Experimentally, this has been demonstrated for many systems \cite{garca,shlyk,li,nishioka,huo,vollmer,sullow2,kalvius,tien}, with theoretical support in the Refs. \cite{theumann,kiselev}.

Specifically, for URh$_2$Ge$_2$ it has been shown that by varying the level of crystallographic disorder a transition from an antiferromagnetic to a spin glass ground state can be facilitated \cite{sullow2}. Further, in the alloying series U(Cu,Pd)$_5$ and (Y,U)Pd$_3$ spin glass ground states are suppressed to zero temperature, and are associated with NFL characteristics \cite{gajewski,vollmer}. Consequently, if glassy states occur in a given class of compounds, we expect a behavior as sketched in Fig. \ref{fig:fig1}(b). Here, for zero disorder upon variation of $J$ the situation is the same as for the purely antiferromagnetic case (Fig. \ref{fig:fig1}(a)). However, if for an antiferromagnetic material the level of disorder is increased from zero, it initially will cause a suppression of $T_N$. Beyond a material specific disorder level a transition into a spin glass ground state below the freezing temperature $T_f$ will occur. Increasing the disorder level further causes a suppression of $T_f$, and the subsequent occurrence of NFL behavior adjacent to the spin glass phase.

In addition to the issues discussed for the purely antiferromagnetic case in Fig. \ref{fig:fig1}(a), with the presence of the spin glass phase additional topics emerge. In particular, it is an open question how spin glass and antiferromagnetic phase compete close to the magnetic instability on the zero-disorder axis. The scenario sketched in Fig. \ref{fig:fig1}(b), with the spin glass phase reaching out to zero disorder, is based on the hypothesis that for an arbitrarily weak magnet an arbitrarily small level of disorder should be sufficient to destroy the AFM state.

For real materials, the situations encountered experimentally might be far more complicated than those that can be derived from the sketches in Fig. \ref{fig:fig1}. We have not considered features like Fermi surface nesting or hidden order phenomena \cite{gu,matsuda,jaime}, nor disorder induced effects like AFM states with anomalously short correlations lengths \cite{tran,maksimov}. Still, from our sketches, the relevance of disorder effects for the understanding and interpretation of experimental data is evident.

So far, we have only addressed the problem of the magnetic ground state properties. Another aspect concerning disorder is its effect on the electronic transport properties. Commonly, for heavy fermions it is assumed that the resistivity $\rho$ probes the magnetic ground state. Correspondingly, anomalous temperature dependencies of $\rho$ in various moderately disordered uranium heavy-fermion compounds have been interpreted in terms of NFL behavior \cite{stewart}.

Recently, based on a detailed study of the electronic transport properties of URh$_2$Ge$_2$, this view has been challenged \cite{sullow3}. Instead, such behavior has now been attributed mostly to disorder induced localization effects. Moreover, for URh$_2$Ge$_2$ the temperature dependence of the anomalous contribution to the Hall effect could not be reconciled with the model commonly used for heavy fermions, e.g., the skew scattering scenario \cite{fert}. Hence, at this point the electronic transport properties of moderately disordered uranium heavy-fermion compounds are neither well studied nor well understood. Therefore, we decided to carry out a detailed investigation of the electronic transport properties of a moderately disordered heavy-fermion compound, in order to provide a data basis for theoretical studies.

For our study we have chosen UCu$_4$Pd, because it is a moderately disordered U heavy fermion which has been characterized in most detail regarding its structural and magnetic properties. Based upon a thermodynamic investigation this system has been established as one of the first NFL materials \cite{andraka}, with the NFL state thoroughly studied by various techniques \cite{bernal,vollmer,aronson}. Initially, the material was thought to crystallize in an ordered lattice derived from the cubic AuBe$_5$ structure, with U, Cu and Pd each occupying separate and translationally invariant sublattices (space group $\overline{43}m$) \cite{chau1}. However, in an XAFS studies it has been demonstrated that the material in fact is disordered, with  25 \% Pd on the nominal Cu sites, and vice versa \cite{booth}. The actual level of disorder depends on the metallurgical treatment of the material, and can be reduced through annealing \cite{booth,weber}.

The resistivity $\rho$ of as-cast UCu$_4$Pd \cite{andraka,weber,chau2} closely resembles the behavior observed for URh$_2$Ge$_2$ \cite{sullow2,sullow3,sullow5}. Absolute values of $\rho$ are large (a few hundred $\mu \Omega cm$), with a negative derivative $d \rho / d T$ down to lowest temperatures. A dependence of $\rho = \rho_0 - A T$ at low $T$ has been attributed to the NFL state. However, annealing the material causes qualitative changes to the resistivity \cite{weber}, as $d \rho / d T$ becomes positive between $\sim 30$ and 150 K, while a dependence $\rho - \rho_0 \propto T$ is no longer observed.

As yet, the mechanism that triggers the transition from a quasi-insulating to a metallic-like state in the resistivity of UCu$_4$Pd upon annealing is not understood. Moreover, the magnetotransport as a tool to distinguish between magnetic and nonmagnetic scattering components has not been investigated in detail. Finally, annealing induced modifications to the band structure like a closing of a pseudogap have not been considered so far. Therefore, here we present a detailed study of the electronic transport properties of as-cast and annealed UCu$_4$Pd, in order to determine if the resistivity probes the NFL state or if it reflects disorder induced localization in a strongly correlated electron material.

\section{Results}

For our study we use data taken on three different sets of samples UCu$_4$Pd, which were provided by the groups in San Diego, Augsburg and Leiden, and are labeled as \textbf{S} (San Diego), \textbf{A} (Augsburg) and \textbf{L} (Leiden), respectively. The samples were prepared under similar conditions, by arc-melting the constituents in stoichiometric ratio in a high-purity argon atmosphere on a water-cooled copper crucible. To improve homogeneity, the samples were remelted and flipped over after each melting process several times. Subsequently, three of the samples have been annealed under various conditions. Altogether, the samples investigated in this paper are as-cast: {\bf S$_{ac}$}, {\bf A$_{ac}$} and {\bf L$_{ac}$}; annealed in an evacuated quartz glass tube at 750$^{\circ}$C for 7 days: {\bf A$_{1ann}$}; at 900$^{\circ}$C for 7 days: {\bf L$_{ann}$}; and at 750$^{\circ}$C for 14 days: {\bf A$_{2ann}$}.

\begin{figure}[!ht]
\begin{center}
\includegraphics[width=1\linewidth]{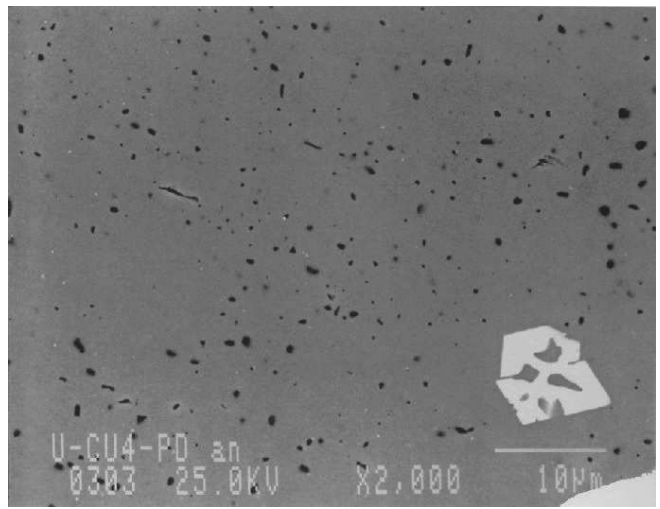}
\end{center}
\caption{Electron--backscattering photograph of the annealed sample UCu$_4$Pd, {\bf L$_{ann}$}. Note the 10 $\mu m$ scale given at the lower right.} \label{fig:fig2}
\end{figure}

Crystallographic structure and homogeneity of the samples \textbf{S} and \textbf{A} have been established by x-ray diffraction. In agreement with previous studies, the samples have been found to crystallize in a AuBe$_5$-type structure. Moreover, no impurity phases could be detected (for details see Refs. \cite{weber,chau2}). In addition, for the samples \textbf{L} secondary phases were not detected in x-ray diffraction. However, electron probe micro analysis (EPMA) carried out on {\bf L$_{ac}$} and {\bf L$_{ann}$} reveals the presence of small inclusions of secondary phases. As a typical example an electron-backscattering photograph of {\bf L$_{ann}$} is displayed in Fig. \ref{fig:fig2}. The black spots and white crystals represent areas with a composition deviating from the nominal 1:4:1. In the backscattering mode regions with a relatively higher concentration of a heavy element appear lighter. Therefore, the black spots are U depleted, while the white crystal is rich in uranium.

In Table \ref{tab:tab1} we summarize the measured compositions and volume amounts of the different metallurgical phases for the samples \textbf{L}. The size of the black inclusions is similar to the sample volume probed by the EPMA electron beam (diameter $\sim 1 \mu m$; see Fig. \ref{fig:fig2}) \cite{sullow6}. Hence, in the analysis part of the surrounding matrix will contribute to the measured signal in a way that the U and Pd compositions are overestimated. Consequently, the black spots (2nd phase) likely are inclusions of pure copper. The white crystal (3rd phase) we believe to be a uranium oxide, which is very stable and, since the light element oxygen is difficult to be measured in EPMA in the presence of heavy elements, would appear as uranium rich area.

\begin{table}
\begin{tabular}{|l|c|c|c|} \hline
Sample & Matrix & 2nd & 3rd \\
& Vol. \% & Vol. \% & Vol. \% \\ \hline
{\bf L$_{ac}$} & 1:4.10(12):1.09(3) & 0.12:1:0.14 & 1:0.11:0.01 \\
& 94 & 5 & 1 \\ \hline
{\bf L$_{ann}$} & 1:3.94(12):1.05(3) & 0.04:1:0.06 & 1:0.06:0.01 \\
& 96 & 3 & 1 \\ \hline
\end{tabular}
\caption{The result of the electron probe micro analysis of the samples UCu$_4$Pd, {\bf L$_{ac}$} and {\bf L$_{ann}$}. Compositions of matrix, 2nd and 3rd phase are given as U:Cu:Pd; for details see text.} \label{tab:tab1}
\end{table}

We have characterized the magnetic ground state and electronic transport properties of our samples by means of susceptibility and resistivity measurements. The transport measurements have been performed using a common four probe technique at temperatures 2 to 300 K (0.03 to 300 K for {\bf A$_{ac}$}, {\bf A$_{2ann}$}). For all samples we present the temperature dependence of the resistivity $\rho (T)$. Some of the data have been published previously \cite{weber,chau2} and are included here for comparison. Also, for the samples {\bf L$_{ac}$} and {\bf L$_{ann}$} we measured the magnetoresistivity $\Delta \rho / \rho = \frac{\rho (B) - \rho(B=0)}{\rho (B=0)}$, the Hall constant $R_H (T)$, and the susceptibility $\chi$. For the magnetotransport and Hall effect measurements the magnetic field $B$ up to 4 T was applied perpendicular to the current direction. The Hall resistance was determined from data taken up to 1.5 T. In this field range we observe a linear dependence of the Hall voltage on $B$.

\begin{figure}[!ht]
\begin{center}
\includegraphics[width=1\linewidth]{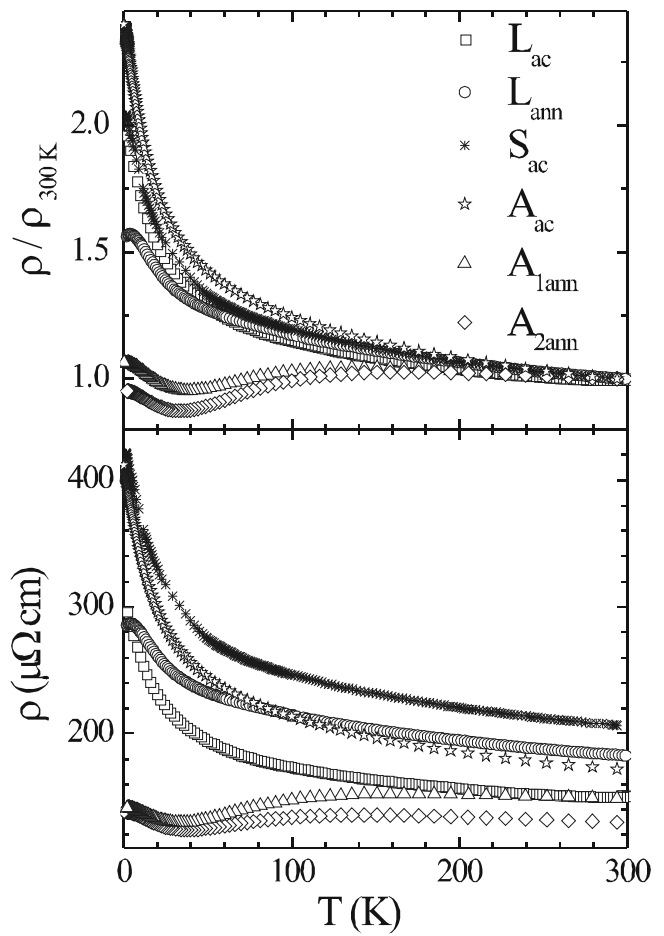}
\end{center}
\caption{The resistivity of as-cast and annealed samples UCu$_4$Pd as a function of temperature, plotted in absolute units and as normalized data.} \label{fig:fig3}
\end{figure}

Figure \ref{fig:fig3} summarizes the temperature dependence of the resistivity of our samples UCu$_4$Pd. For as-cast material {\bf S$_{ac}$}, {\bf A$_{ac}$} and {\bf L$_{ac}$} we observe the archetypical behavior of moderately disordered uranium heavy-fermion compounds, with a negative $d \rho / d T$ and large absolute $\rho$ values. Notably, the as-cast material $\rho$ values differ by about $100 \mu \Omega cm$, while the normalized resistivity exhibits a very similar behavior for the three samples (see Tab. \ref{tab:tab2}). This sample-to-sample variation of $\rho$ indicates that structural disorder affects the transport behavior.

\begin{table}
\begin{tabular}{|l|c|c|c|} \hline
Sample & $\rho_0 (\mu \Omega cm)$ & $T_0$ (K) & $T_1$ (K) \\ \hline
{\bf S$_{ac}$} & 429 & 74 & 14 \\ \hline
{\bf L$_{ac}$} & 304 & 68 & 12 \\ \hline
{\bf A$_{ac}$} & 412 & 55 & 8 \\ \hline
Ref.\cite{chau2} & 258 & 64 & 10 \\\hline
\end{tabular}
\caption{The parameters resulting from a fit of the low temperature resistivity $\rho (T ) = \rho_0 (1 - T/T_0)$ of as-cast UCu$_4$Pd; for details see text.} \label{tab:tab2}
\end{table}

\begin{figure}[!ht]
\begin{center}
\includegraphics[width=1\linewidth]{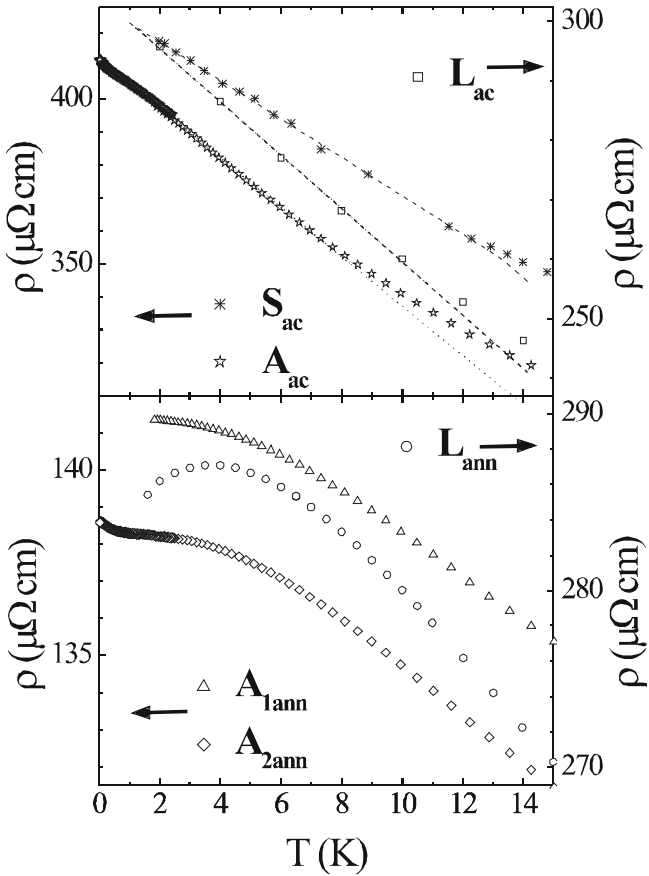}
\end{center}
\caption{The low temperature resistivity of as-cast and annealed samples UCu$_4$Pd as a function of temperature.} \label{fig:fig4}
\end{figure}

At low temperatures, as has been stated previously \cite{andraka}, the resistivity is linear in temperature (Fig. \ref{fig:fig4}). Following Ref. \cite{chau2} we fit the resistivity as $\rho (T ) = \rho_0 (1 - T/T_0)$. The resulting fit parameters are summarized in Tab. \ref{tab:tab2}. Here, $T_1$ denotes the temperature up to which the data are fitted. Again, we find significant sample-to-sample variations indicative of a varying level of disorder.

Annealing the material qualitatively and quantitatively changes the resistive behavior \cite{weber}. Comparing {\bf L$_{ann}$}, {\bf A$_{1ann}$} and {\bf A$_{2ann}$} we find pronounced sample-to-sample variations. This implies that for annealed material disorder still affects the resistive behavior. Conversely, concerning the overall temperature dependence the annealed samples show a more metallic behavior than the as-cast ones. In particular, for the samples {\bf A$_{1ann}$} and {\bf A$_{2ann}$} a metallic (i.e., positive) $d \rho / d T$ occurs for intermediate temperatures, and $\rho$ at lowest $T$ saturates. For {\bf L$_{ann}$}, while the absolute value is surprisingly larger than that of the as-cast sample {\bf L$_{ac}$}, the resistivity ratio $\rho_{2K} / \rho_{300K}$ is smaller or more metallic, and at the lowest $T$ we observe a shallow maximum $\sim 4$ K in $\rho$.

\begin{figure}[!ht]
\begin{center}
\includegraphics[width=1\linewidth]{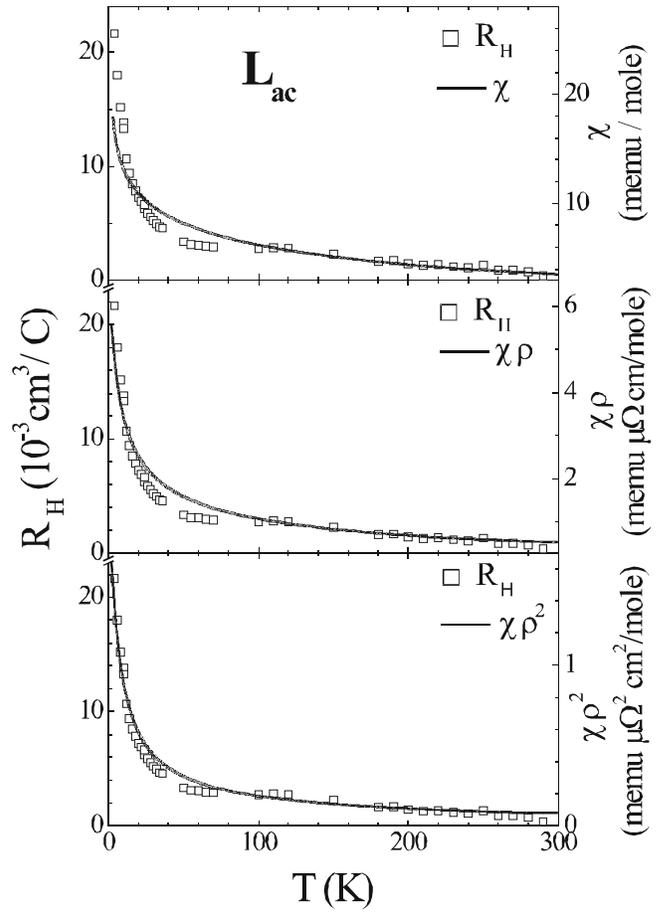}
\end{center}
\caption{The Hall constant of UCu$_4$Pd, {\bf L$_{ac}$}, as a function of temperature. We include the temperature dependence of the susceptibility $\chi$ and of the products $\chi \rho$, $\chi \rho^2$; for details see text.} \label{fig:fig5}
\end{figure}

Since for the annealed sample {\bf L$_{ann}$} we find no anomaly in the $T$ dependence of the susceptibility (see below Fig. \ref{fig:fig5}) in the temperature range of the resistive maximum, we conclude that the maximum does not correspond to a magnetic phase transition. Instead, it could stem either from regions in our material, in which disorder has been minimized to such a degree that a coherent state can be formed, or from a grain boundary phase. We note that for a grain boundary phase there is no metallurgical evidence, as in the EPMA backscattering photo we do not see it. Rather, the topological distribution of secondary phases in the annealed sample is similar to the as-cast
one, in which no resistive anomaly was observed.

\begin{figure}[!ht]
\begin{center}
\includegraphics[width=1\linewidth]{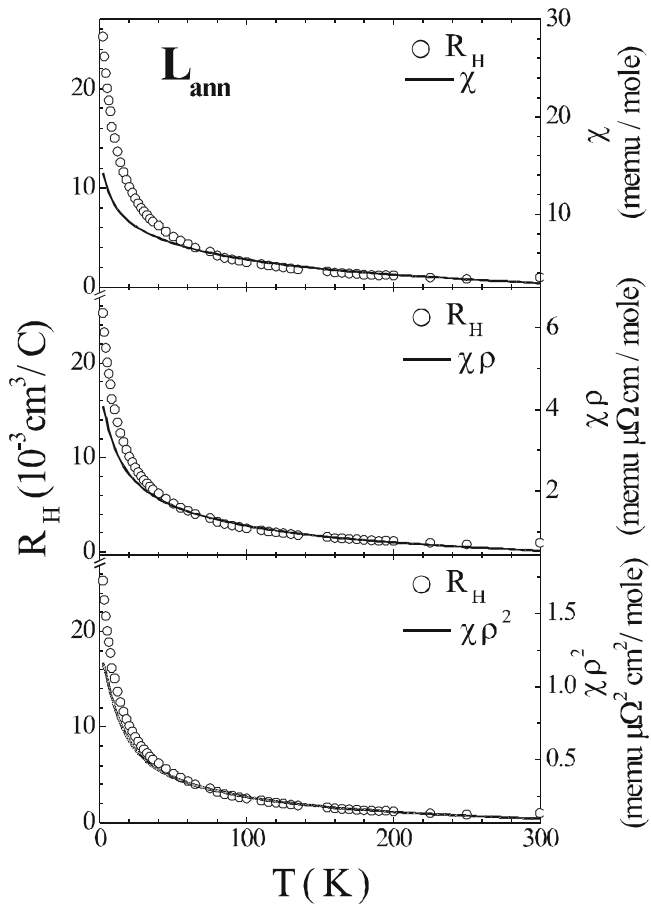}
\end{center}
\caption{The Hall constant of UCu$_4$Pd, {\bf L$_{ann}$}, as a function of temperature. We include the temperature dependence of the susceptibility $\chi$ and of the products $\chi \rho$, $\chi \rho^2$; for details see text.} \label{fig:fig6}
\end{figure}

To establish the carrier density for UCu$_4$Pd we have carried out Hall effect measurements on {\bf L$_{ac}$} and {\bf L$_{ann}$}. In the Figs. \ref{fig:fig5} and \ref{fig:fig6} we plot the temperature dependence of the Hall constant $R_H$ for both samples. Overall, these $R_H (T)$ for {\bf L$_{ac}$} and {\bf L$_{ann}$} are very similar. However, they are unlike what is expected for archetypical heavy fermion metals \cite{fert,kontani}. In general, for heavy fermions a maximum in the $T$ dependence of the Hall constant, denoting the transition into the low temperature coherent state below $T_{coh}$, is observed. In UCu$_4$Pd there is no such maximum, reflecting the absence of coherent scattering in this disordered compound, i.e., no coherent state.

The strong $T$ dependence of $R_H$ indicates that it is dominated by anomalous contributions. As usual, for the analysis of the Hall effect data we assume a temperature independent carrier density and a spherical Fermi surface. To extract the carrier density we parameterize the Hall constant as
\begin{equation}
R_H = R_0 + \chi R_{S,mag}. \label{eq:eq1}
\end{equation}
Here, $R_0$ denotes the normal contribution to the Hall effect and measures the carrier density $n$. The second term, $\chi R_{S,mag}$, represents the anomalous contribution to the Hall effect, with $R_{S,mag}$ a model dependent factor. For heavy fermions, in the first Hall effect studies an empirical ansatz had been made that $R_{S,mag} = const$ \cite{schoenes}. In contrast, the skew scattering scenario \cite{fert}, which is characterized by a constant spontaneous angle $\alpha$ at which the scattered carriers are deflected from their original trajectories, this as result of spin-orbit scattering, predicts that $R_{S,mag} \propto \rho_{mag}$, with $\rho_{mag}$ as the magnetic scattering component of the resistivity. For heavy fermion materials, at high temperatures $\rho_{mag}$ is essentially constant, and it goes to zero for $T \rightarrow 0$. Hence, the product $\chi \rho_{mag}$ will exhibit a pronounced maximum at $T_{coh}$. Alternatively, the coherence maximum in $R_H$ has been explained on basis of the periodic Anderson model \cite{kontani} as a transition from a regime $T \ll T_{coh}$ with $R_{S,mag} \propto \rho^2 / \chi$ to one at $T \gg T_{coh}$, $R_{S,mag} = const$, with $\rho$ now the total resistivity. Finally, for disordered materials the side jump effect has been identified in Hall effect measurements \cite{berger}, giving rise to a $T$ dependence $R_{S,mag} \propto \rho^2$. The side jump mechanism is quantum mechanical in nature and results in a constant lateral displacement $\Delta y$ of the charge trajectory at the point of scattering.

To assess which of these models most adequately describes our data we include the measured susceptibility $\chi (T)$ and the products $\chi \rho$, $\chi \rho^2$ in the Figs. \ref{fig:fig5} and \ref{fig:fig6}. All in all, from our comparison we find that at high temperatures our Hall effect data are reproduced by all three scenarios. This simply reflects the comparatively weak $T$ dependence of $\rho$. Then, by extracting the normal Hall effect contribution $R_0 = (ne)^{-1}$, for both samples we obtain a carrier density $n \sim 5 \times 10^{21}$ carriers/cm$^3$, corresponding to 1 to 2 electrons per unit cell, irrespective of the assumptions made for the anomalous contribution. This carrier density is metallic and disproves the possibility for a gap in the density of states. Furthermore, the annealing induced qualitative changes of the resistivity cannot be associated to the carrier density.

For the overall $T$ dependence of $R_H$ we observe that the fitting is best for the product $\chi \rho^2$, the side jump effect scenario \cite{note2}. Such might indicate a disorder dominated Hall effect, since usually the side jump effect is observed in disordered materials like metallic glasses \cite{mayeya}.

In contrast and strictly speaking, we should have plotted $\rho_{mag}$ for the skew scattering scenario. However, for the anomalous resistivities of as-cast and annealed UCu$_4$Pd a correction for phonon contributions - as it is usually performed - will become an ad hoc procedure. Therefore, we use the total resistivity, assuming that it approximates $\rho_{mag}$. In this way, we implicitely assume a temperature dependence of the magnetic resistive component unlike that used in Ref. \cite{fert}, since $\rho_{mag}$ for $T \rightarrow 0$ does not approach zero. To the best of our knowledge, aside from hand waving arguments presented in context with Kondo disorder scenarios \cite{miranda,castro}, there is no well-founded model that would explain such a highly anomalous magnetic resistive component. This inconsistency would imply that the skew scattering scenario, as set out by Fert and Levy \cite{fert}, does not account for the anomalous Hall contribution observed for UCu$_4$Pd.

\begin{figure}[!ht]
\begin{center}
\includegraphics[width=1\linewidth]{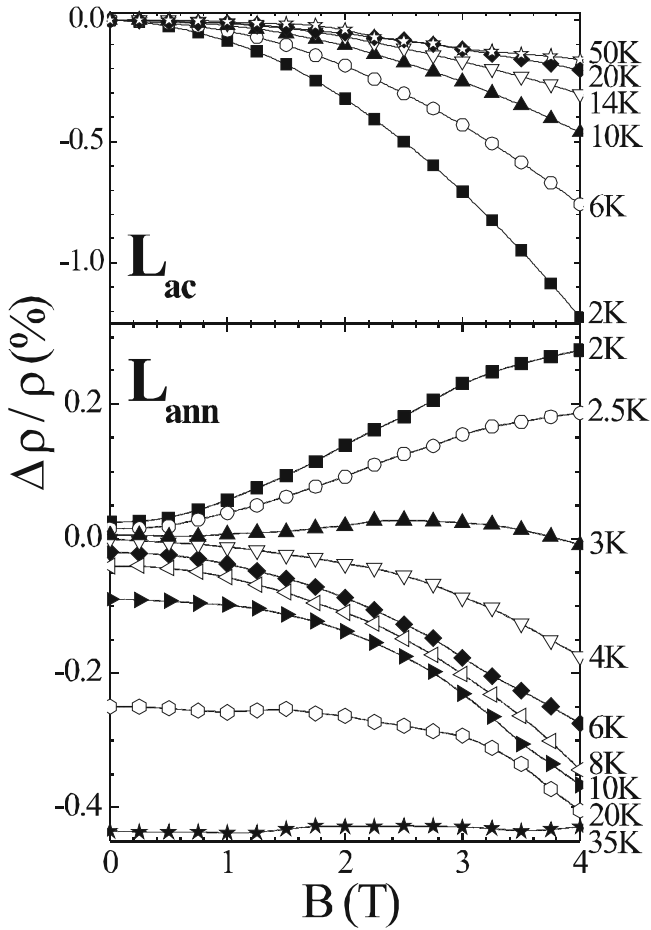}
\end{center}
\caption{The magnetoresistivity of the annealed sample UCu4Pd, {\bf L$_{ann}$}; data are offset for clarity. For comparison, we include data taken on the as-cast sample {\bf L$_{ac}$}; for details see text.} \label{fig:fig7}
\end{figure}

To further study the magnetic resistive component we have carried out magnetoresistivity measurements on {\bf L$_{ac}$} and {\bf L$_{ann}$}. The naive expectation would be that - if the low temperature resistivity is mostly magnetic in nature - it must be spin disorder scattering, which should be substantially reduced in magnetic fields of a few T. Correspondingly, in Fig. \ref{fig:fig7} we plot the magnetoresistivity at various temperatures for the annealed sample {\bf L$_{ann}$}. For comparison, we include data taken on {\bf L$_{ac}$}. A full account of the measurements on the as-cast sample has been given elsewhere \cite{otop}, where we have demonstrated that for the as-cast sample at temperatures below 30 K the field dependence up to highest fields is well described by
\begin{equation}
\Delta \rho / \rho = \alpha (T) B^2. \label{eq:eq2}
\end{equation}
This, and the fact that $\alpha (T)$ follows the susceptibility ($\alpha \propto \chi - \chi_0$), indicates that spin disorder scattering is the dominant process controlling the field dependence of $\Delta \rho / \rho$ for {\bf L$_{ac}$}. For {\bf L$_{ann}$}, however, we observe additional structure in $\Delta \rho / \rho$, prohibiting an analysis using Eq. \ref{eq:eq2}. Moreover, while $\alpha$ for {\bf L$_{ac}$} is a monotonous function of temperature, for {\bf L$_{ann}$} the magnetoresistivity changes sign from positive to negative at around 3 K.

\begin{figure}[!ht]
\begin{center}
\includegraphics[width=1\linewidth]{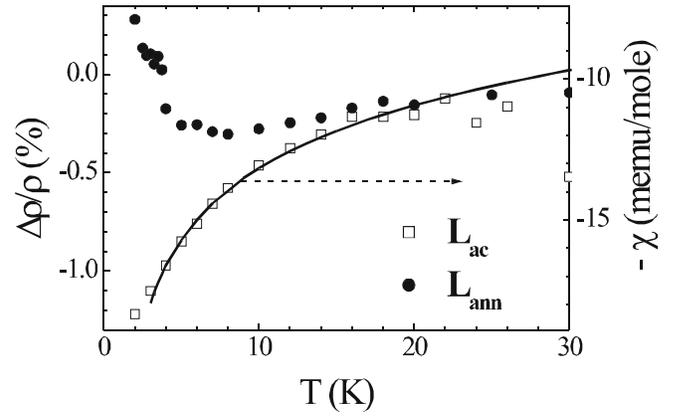}
\end{center}
\caption{The temperature dependence of the magnetoresistivity $\Delta \rho / \rho$ for the as-cast and annealed samples {\bf L$_{ac}$} and {\bf L$_{ann}$}, determined in an external field of 4 T. We include the magnetic susceptibility of the as-cast sample (solid line - right scale); for details see text.} \label{fig:fig8}
\end{figure}

To quantitatively compare the magnetoresistivities of {\bf L$_{ac}$} and {\bf L$_{ann}$}, in Fig. \ref{fig:fig8} we plot the values $\Delta \rho / \rho$ determined in an external field of 4 T. Overall, the size of $\Delta \rho / \rho$ is small, being less than 1.3 \% and 0.4 \% for {\bf L$_{ac}$} and {\bf L$_{ann}$}, respectively. This alone suggests that the large values of the resistivity do not arise from magnetic scattering. It thus represents additional evidence against an interpretation of the anomalous Hall contribution in terms of skew scattering theory \cite{fert}.

Again, we can compare the magnetoresistivity with the magnetic response of our samples, that is, the susceptibility. As pointed out above, for {\bf L$_{ac}$} we find $\Delta \rho / \rho \propto (\chi - \chi_0)$ consistent with spin disorder scattering. For {\bf L$_{ann}$}, however, the behavior is more complex. While above $\sim 15$ K the $T$ dependence of $\Delta \rho / \rho$ coincides with that of {\bf L$_{ac}$}, at low temperatures the magnetoresistivity changes sign. In the resistivity this feature is associated to the broad maximum at around 4 K (see Fig. \ref{fig:fig4}). Since the maximum likely does not represent a bulk phenomenon, we believe that the change of sign of the magnetoresistivity for {\bf L$_{ann}$} as well does not constitute bulk behavior. We will briefly address this point later in the discussion.

\section{Discussion}

With our data analysis we have established the metallicity of both as-cast and annealed UCu$_4$Pd. Then, the negative $d \rho / d T$ for the as-cast material must be the result from disorder-induced localization effects. Further, the electronic transport properties of UCu$_4$Pd closely resemble those of the moderately disordered heavy-fermion compound URh$_2$Ge$_2$, which again are reminiscent of metallic glasses \cite{sullow3}. For the latter compound the metallic glass like behavior has been quantified by applying the corresponding localization theory to describe the $T$ dependence of the conductivity $\sigma (T)$.

\begin{figure}[!ht]
\begin{center}
\includegraphics[width=1\linewidth]{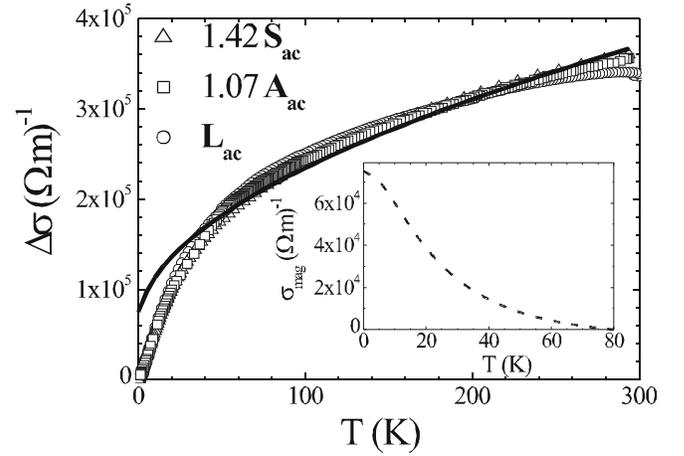}
\end{center}
\caption{The temperature dependence of the reduced conductivity $\Delta \sigma$ for as-cast UCu$_4$Pd. The data for {\bf S$_{ac}$} and {\bf A$_{ac}$} have been scaled onto those of {\bf L$_{ac}$} by multiplying them with a constant factor of 1.42 and 1.07, respectively. The solid line depicts the result of a parametrization of the data. In the inset we plot the low $T$ difference between the parametrization and the experimental data for the sample {\bf A$_{ac}$}; for details see text.} \label{fig:fig9}
\end{figure}

Following the procedure set out in Ref. \cite{sullow3}, in Fig. \ref{fig:fig9} we plot the reduced conductivity $\Delta \sigma = \sigma - \sigma (T = 0)$ as a function of $T$. By multiplying the data for {\bf S$_{ac}$} and {\bf A$_{ac}$} with a constant factor of 1.42 and 1.07, respectively, we can scale these conductivity data onto those of {\bf L$_{ac}$} over the complete temperature range, with only some minor mismatch at highest $T$ between ({\bf S$_{ac}$}, {\bf A$_{ac}$}) and {\bf L$_{ac}$}. It illustrates the close similarity of the $T$ dependence of the electronic transport properties for the different as-cast samples UCu$_4$Pd.

To quantify the resemblance to metallic glasses we describe our data in terms of localization theory \cite{sullow3,lee}. For weak electronic correlations (in UCu$_4$Pd at sufficiently high $T$) the $T$ dependence of $\sigma$ is attributed to the superposition of incipient localization, destroyed by inelastic scattering with phonons and electrons, and electronic interaction effects. It is given by \cite{hickey}
\begin{equation}
\Delta \sigma = \frac{e^2}{2 \pi^2 \hbar} \left( 3 \sqrt{b + c^2 T^2} - cT - 3 \sqrt{b} + d \sqrt{T} \right), \label{eq:eq3}
\end{equation}
with fit parameters $b = 1 / D \tau_{so}$, $c = \sqrt{1 / 4 D \beta}$, $\beta = \tau_i T^2$, $d = 0.7367 \sqrt{k / D \hbar}$ (diffusion coefficient $D$, spin-orbit ($\tau_{so}$) and inelastic ($\tau_i$) scattering times) \cite{note4}. Quantitatively, above 60 K the $T$ dependence of $\Delta \sigma$ can nicely be described by Eq. \ref{eq:eq3}, validating our statement on the close resemblance to the behavior of amorphous metals. In Fig. \ref{fig:fig9} we include the result of a parametrization of the data for sample {\bf A$_{ac}$} as a solid line. Because of parameter interdependencies we cannot extract a unique and meaningful set of fit parameters \cite{note3}. Overall, the essential features of the data for the three samples can be reproduced using parameters $D \sim 10^{-7} - 10^{-8}$m$^2$/s; $\tau_{so} \sim 10^{-9} - 10^{-10}$s and $\beta \sim 10^{-7} - 10^{-9}$sK$^2$, which are similar in order of magnitude to those obtained for URh$_2$Ge$_2$.

While there are still open questions as to the applicability of classical localization theory to the transport properties of a moderately disordered uranium heavy fermion compound, our analysis indicates that the behavior of our compound at high enough temperatures is metallic-glass like. In particular, with Matthiesen's rule not being valid, it implicates that the transport properties of moderately disordered uranium heavy fermion compounds should be considered in terms of transport through different conductivity channels.

If we proceed in the spirit of localization theory, we can attribute the difference between our parametrization and the experimental data at low temperatures to a conductivity channel which contains a magnetic component. The idea is that at high temperatures this channel fully conducts, but at low temperatures its conductivity is reduced because of magnetic scattering. The reduction of the conductivity of this channel we can determine from $\sigma_{mag} = \Delta \sigma - \sigma_{exp}$, with the data for sample {\bf A$_{ac}$} plotted in the inset of Fig. \ref{fig:fig9}. Due to the uncertainty of the fitting parameters in $\Delta \sigma$ our analysis procedure can only estimate the relevance of different transport mechanisms. The value of $\sigma_{mag} (T \rightarrow 0)$, $7 \times 10^4 (\Omega \rm m)^{-1}$, corresponds to about 30 \% of the total conductivity of this sample at low temperatures. This is a much more realistic estimate of the size of the magnetic contribution to the electronic transport than by assuming that the total resistivity reflects the magnetic scattering. Moreover, it should be such a ''magnetic conductivity channel'' which captures the NFL characteristics from quantum spin fluctuations rather than the $T$ dependence of $\rho$ at low temperatures \cite{stewart}. Unfortunately, at present there are no theories available which would allow a more detailed analysis.

Finally, the pronounced sample dependence of the electronic transport properties ought to be addressed. We have demonstrated that aside from differences in absolute values of the conductivity the $T$ dependencies for the as-cast samples are essentially the same. This could be explained by assuming that the effective path of the electrical current is different from sample to sample. It would imply that the current in as-cast samples UCu$_4$Pd follows a percolative path provided by the better ordered regions in the sample. Along this path transport is diffusive and exhibits the characteristic negative $d \rho / d T$ of moderately disordered uranium heavy fermions. If by annealing, in some part of this percolative path, the disorder is reduced to a degree that transport starts to become ballistic, the superposition of ballistic and diffusive transport will yield a temperature dependence of $\rho$ similar to what is observed experimentally. Here, in order to test such a scenario local probe experiments of the electronic transport, like scanning tunneling microscopy or point contact spectroscopy might be useful.

In conclusion, we have presented a detailed study of the electronic transport properties of the moderately disordered uranium heavy fermion UCu$_4$Pd. With our study we have established the metallicity of the material for both as-cast and annealed material. Our data analysis indicates that the electronic transport properties are dominated by disorder-induced localization effects, while there is only a minor magnetic scattering component. This observation casts doubt on claims for NFL behavior detected in the resistivity of this and related compounds.

\section{Acknowledgements}

This work has been supported by the High Magnetic Field Laboratory Braunschweig and by the Deutsche Forschungsgemeinschaft DFG under contract no. SU229/1-3 and through SFB 484. Research at UCSD was supported by the US Department of Energy under Grant No. FG02-04ER46105. In part, samples have been produced within FOM/ALMOS.


\begin{references}
\bibitem{dalichaouch} Y. Dalichaouch, M.C. de Andrade, D.A. Gajewski, R. Chau, P. Visani, and M.B. Maple, Phys. Rev. Lett. \textbf{75}, 3938 (1995).
\bibitem{sullow1} S. S\"ullow, B. Ludoph, B. Becker, G. J. Nieuwenhuys, A.A. Menovsky, and J.A. Mydosh, Phys. Rev. B \textbf{56}, 846 (1997).
\bibitem{joynt} R. Joynt and L. Taillefer, Rev. Mod. Phys. \textbf{74}, 235 (2002).
\bibitem{zapf} V.S. Zapf, E.J. Freeman, E.D. Bauer, J. Petricka, C. Sirvent, N.A. Frederick, R.P. Dickey, and M.B. Maple, Phys. Rev. B \textbf{65}, 014506 (2002).
\bibitem{miranda} E. Miranda, V. Dobrosavljevic, and G. Kotliar, Phys. Rev. Lett. \textbf{78}, 290 (1997); E. Miranda and V. Dobrosavljevic, Phys. Rev. Lett. \textbf{86}, 264 (2001).
\bibitem{andrade} M.C. de Andrade, R. Chau, R.P. Dickey, N.R. Dilley, E.J. Freeman, D.A. Gajewski, M.B. Maple, R. Movshovich, A.H. Castro Neto, G. Castilla, and B.A. Jones, Phys. Rev. Lett. \textbf{81}, 5620 (1998).
\bibitem{castro} A.H. Castro Neto, G. Castilla, and B.A. Jones, Phys. Rev. Lett. \textbf{81}, 3531 (1998); A.H. Castro Neto and B.A. Jones, Phys. Rev. B \textbf{62}, 14975 (2000).
\bibitem{rosch} A. Rosch, Phys. Rev. Lett. \textbf{82}, 4280 (1999); Phys. Rev. B \textbf{62}, 4945 (2000).
\bibitem{wysokinski} K.I. Wysokinski, Phys. Rev. B \textbf{60}, 16376 (1999).
\bibitem{garca} J. Garcia Soldevilla, J.C. G\'omez Sal, J.A. Blanco, J.I. Espeso, and J. Rodriguez Fern\'andez, Phys. Rev. B \textbf{61}, 6821 (2000).
\bibitem{shlyk} L. Shlyk and J. Stepien-Damm, J. Magn. Magn. Mater. \textbf{195}, 37 (1999); F.G. Gandra, D.P. Rojas, L. Shlyk, L.P. Cardoso, and A.N. Medina, J. Magn. Magn. Mat. \textbf{226-230}, 1312 (2001).
\bibitem{pechev} S. Pechev, T. Roisnel, B. Chevalier, B. Darriel, and J. Etourneau, Solid State Sci. \textbf{2}, 773 (2000).
\bibitem{gajewski} D.A. Gajewski, R. Chau, and M.B. Maple, Phys. Rev. B \textbf{62}, 5496 (2000).
\bibitem{li} D.X. Li {\it et al.}, J. Magn. Magn. Mater. 176, 261 (1997); Phys. Rev. B \textbf{57}, 7434 (1998); Solid State Commun. \textbf{108}, 163 (1998); J. Phys.: Condens. Matter \textbf{11}, 8263 (1999).
\bibitem{nishioka} T. Nishioka, Y. Tabata, T. Taniguchi, and Y. Miyako, J. Phys. Soc. Jpn. \textbf{69}, 1012 (2000).
\bibitem{huo} D. Huo, J. Sakurai, T. Kuwai, Y. Isikawa, and Q. Lu, Phys. Rev. B \textbf{64}, 224405 (2001).
\bibitem{tran} V.H. Tran, F. Steglich, and G. Andr\'e, Phys. Rev. B \textbf{65}, 134401 (2002).
\bibitem{bernal} O.O. Bernal, D.E. MacLaughlin, H.G. Lukefahr, and B. Andraka, Phys. Rev. Lett. \textbf{75}, 2023 (1995); D.E. MacLaughlin, O.O. Bernal, R.H. Heffner, G.J. Nieuwenhuys, M.S. Rose, J.E. Sonier, B. Andraka, R. Chau, and M.B. Maple, Phys. Rev. Lett. \textbf{87}, 066402 (2001).
\bibitem{vollmer} R. Vollmer, T. Pietrus, H.v. L\"ohneysen, R. Chau, and M.B. Maple, Phys. Rev. B \textbf{61}, 1218 (2000).
\bibitem{aronson} M.C. Aronson, R. Osborn, R. Chau, M.B. Maple, B.D. Rainford, and A.P. Murani, Phys. Rev. Lett. \textbf{87}, 197205 (2001).
\bibitem{booth} C.H. Booth, D.E. MacLaughlin, R.H. Heffner, R. Chau, M.B. Maple, and G.H. Kwei, Phys. Rev. Lett. \textbf{81}, 3960 (1998); E.D. Bauer, C.H. Booth, G.H. Kwei, R. Chau, and M.B. Maple, Phys. Rev. B \textbf{65}, 245114 (2002); C.H. Booth, E.­W. Scheidt, U. Killer, A. Weber, and S. Kehrein, Phys. Rev. B \textbf{66}, 140402(R) (2002).
\bibitem{sullow2} S. S\"ullow, G.J. Nieuwenhuys, A.A. Menovsky, J.A. Mydosh, S.A.M. Mentink, T.E. Mason, and W.J.L. Buyers, Phys. Rev. Lett. \textbf{78}, 354 (1997); S. S\"ullow, S.A.M. Mentink, T.E. Mason, R. Feyerherm, G.J. Nieuwenhuys, A.A. Menovsky, and J.A. Mydosh, Phys. Rev. B \textbf{61}, 8878 (2000).
\bibitem{sullow3} S. S\"ullow, I. Maksimov, A. Otop, F.J. Litterst, A. Perucchi, L. Degiorgi, and J.A. Mydosh, Phys. Rev. Lett. \textbf{93}, 266602 (2004).
\bibitem{graf} T. Graf, J.D. Thompson, M.F. Hundley, R. Movshovich, Z. Fisk, D. Mandrus, R.A. Fisher, and N.E. Phillips, Phys. Rev. Lett. \textbf{78}, 3769 (1997).
\bibitem{kalvius} G.M. Kalvius, K. Kojima, M. Larkin, G.M. Luke, J. Merrin, B. Nachumi, Y.J. Uemura, A. Br\"uckel, K. Neumaier, K. Andres, C. Paulsen, G. Nakamoto, and T. Takabatake, Physica B \textbf{281-282}, 66 (2000).
\bibitem{tien} C. Tien, J.J. Lu, and L.Y. Jang, Phys. Rev. B \textbf{65}, 214416 (2002).
\bibitem{doniach} S. Doniach, in \emph{Valence Instabilities and Related Narrow Band Phenomena}, edited by R. D. Parks (Plenum, New York, 1977), p. 169; Physica B \textbf{91}, 231 (1977).
\bibitem{sullow4} S. S\"ullow, M.C. Aronson, B.D. Rainford, and P. Haen, Phys. Rev. Lett. \textbf{82}, 2963 (1999).
\bibitem{hertz} J.A. Hertz, Phys. Rev. B \textbf{14}, 1165 (1976).
\bibitem{millis} A.J. Millis, Phys. Rev. B \textbf{48}, 7183 (1993).
\bibitem{si} Q. Si, S. Rabello, K. Ingersent, and J.L. Smith, Nature \textbf{413}, 804 (2001); D.R. Grempel and Q. Si, Phys. Rev. Lett. \textbf{91}, 026401 (2003); L. Zhu, M. Garst, A. Rosch, and Q. Si, Phys. Rev. Lett. \textbf{91}, 066404 (2003).
\bibitem{theumann} A. Theumann, B. Coqblin, S.G. Magalhaes, and A.A. Schmidt, Phys. Rev. B \textbf{63}, 054409 (2001).
\bibitem{kiselev} M. Kiselev, K. Kikoin, and R. Oppermann, Phys. Rev. B \textbf{65}, 184410 (2002).
\bibitem{griffiths} R.B. Griffiths, Phys. Rev. Lett. \textbf{23}, 17 (1969), and references therein.
\bibitem{zhou} G.F. Zhou and H. Bakker, Phys. Rev. Lett. \textbf{72}, 2290 (1994); \emph{ibid.} \textbf{73}, 344 (1994).
\bibitem{note1} This is in contrast to large moment magnets and weakly correlated metals, where a complete suppression of the magnetic transition temperatures through disorder appears to be difficult; see Ref. \cite{zhou}.
\bibitem{lapertot} G. Lapertot, R. Calemczuk, C. Marcenat, J.Y. Henry, J.X. Boucherle, J. Flouquet, J. Hammann, R. Cibin, J. Cors, D. Jaccard, and J. Sierro, Physica B \textbf{186-188}, 454 (1993).
\bibitem{mentink} S.A.M. Mentink, G.J. Nieuwenhuys, A.A. Menovsky, J.A. Mydosh, H. Tou, and Y. Kitaoka, Phys. Rev. B \textbf{49}, 15759 (1994).
\bibitem{gu} H. Gu, J. Tang, A. Matsushita, T. Taniguchi, Y. Tabata, and Y. Miyako, Phys. Rev. B \textbf{65}, 024403 (2002).
\bibitem{matsuda} K. Matsuda, Y. Kohori, T. Kohara, K. Kuwahara, and H. Amitsuka, Phys. Rev. Lett. \textbf{87}, 087203 (2001).
\bibitem{jaime} M. Jaime, K.H. Kim, G. Jorge, S. McCall, and J.A. Mydosh, Phys. Rev. Lett. \textbf{89}, 287201 (2002).
\bibitem{maksimov} I. Maksimov, F.J. Litterst, H. Rechenberg, M.A.C. de Melo, R. Feyerherm, R.W.A. Hendrikx, T.J. Gortenmulder, J.A. Mydosh, and S. S\"ullow, Phys. Rev. B \textbf{67}, 104405 (2003).
\bibitem{stewart} G. Stewart, Rev. Mod. Phys. \textbf{73}, 797 (2001).
\bibitem{fert} A. Fert and P.M. Levy, Phys. Rev. B \textbf{36}, 1907 (1987).
\bibitem{andraka} B. Andraka and G.R. Stewart, Phys. Rev. B \textbf{47}, 3208 (1993).
\bibitem{chau1} R. Chau, M.B. Maple, and R.A. Robinson, Phys. Rev. B \textbf{58}, 139 (1998).
\bibitem{weber} A. Weber, S. K\"orner, E.W. Scheidt, S. Kehrein, and G.R. Stewart, Phys. Rev. B \textbf{63}, 205116 (2001).
\bibitem{chau2} R. Chau and M.B. Maple, J.Phys.: Condens. Matter \textbf{8}, 9939 (1996).
\bibitem{sullow5} S. S\"ullow, M.B. Maple, R. Chau, D. Tomuta, G.J. Nieuwenhuys, A.A. Menovsky, and J.A. Mydosh, J. Magn. Magn. Mat. \textbf{226-230}, 35 (2001).
\bibitem{sullow6} S. S\"ullow, T.J. Gortenmulder, G.J. Nieuwenhuys, A.A. Menovsky, and J.A. Mydosh, J. Alloys Comp. \textbf{215}, 223 (1994).
\bibitem{kontani} H. Kontani and K. Yamada, J. Phys. Soc. Jpn. \textbf{63}, 2627 (1994).
\bibitem{schoenes} J. Schoenes, J.J.M. Franse, Phys. Rev. B \textbf{33}, R5138 (1986); J. Schoenes, C. Sch\"onenberger, J.J.M. Franse, and A.A. Menovsky, Phys. Rev. B \textbf{35}, R5375 (1987).
\bibitem{berger} L. Berger, Phys. Rev. B \textbf{11}, 4559 (1970).
\bibitem{note2} In Ref. \cite{kontani} the $\rho^2$ dependence is predicted and observed only at $T \ll  T_{coh}$. With the absence of coherent scattering in UCu$_4$Pd this prediction appears not to applicable here.
\bibitem{mayeya} F.M. Mayeya and M.A. Howson, Phys. Rev. B \textbf{49}, 3167 (1994).
\bibitem{otop} A. Otop, F.J. Litterst, T.J. Gortenmulder, J.A. Mydosh, and S. S˜ullow, Physica B, in print (2005).
\bibitem{lee} P.A. Lee and T.V. Ramakrishnan, Rev. Mod. Phys. \textbf{57}, 287 (1985); B.L. Altshuler and A.G. Aronov, in \emph{Electron­Electron Interaction in Disordered Systems}, edited by A.L. Efros and M. Pollak (Elsevier, New York, 1985), p. 1.
\bibitem{hickey} B.J. Hickey, D. Greig, and M.A. Howson, J. Phys. F: Met. Phys. \textbf{16} (1986) L13; H.H. Boghosian and M.A. Howson, Phys. Rev. B \textbf{41}, 7397 (1990).
\bibitem{note4} $\tau_i$ is controlled by phonon scattering, causing a temperature dependence $\tau_i$\,$\propto$\,$T^{-2}$. Thus, the coefficient $\beta$ is independent of temperature.
\bibitem{note3} The plotted curve has been calculated using parameters of $D \sim 6 \times 10^{-8}$m$^2$/s; $\tau_{so} \sim 2 \times 10^{-9}$s; $\beta \sim 2 \times 10^{-7}$sK$^2$.
\end{references}
\end{document}